# Role of self-propulsion of marine larvae on their probability of contact with a protruding collector located in a sea current

Gregory Zilman[1*], Julia Novak[1], Alex Liberson[1], Shimrit Perkol-Finkel[1], Yehuda Benayahu[2]

1. School of Mechanical Engineering, Tel-Aviv University, 69978, Israel, 2. School of Life Science, Tel- Aviv University, 69978, Israel.

Settlement of marine larvae on a substrate is a fundamental problem of marine life. The probability of settlement is one of the quantitative characteristic of the settlement process. The probability of larval contact with a substrate is the upper bound of the probability of settlement. This work addresses the problem of contact probability and contact rate of marine invertebrate larvae with an isolated protruding collector located in an unbounded sea current. There are two common approaches to the problem of contact probability. In one, a collector induces certain cues, which help a larvae find the collector. In such a case, the larva moves towards the collector deliberately, using its navigation and propulsion devices. In the second approach, a larva moves towards a collector as a passive small particle. In this case, the cause of contact of a larva with a collector is a mechanical collision of a small moving body with a large obstacle. We considered a larva which does not know the location of the collector, which does not use its navigation device yet uses its self-propulsion. We mimiced a larva by a tiny self-propelled underwater vehicle, moving in shear flow of a large obstacle. We illustrated our approach by studying contact of a larva of the Bryozoan *Bugula neritina* with a cylindrical collector. We observed the behavior of this larva in a laboratory flume, and according to the observations formulated a mathematical model of larval motion in shear flow. The trajectories of a large number of larvae, starting their motion far from a collector with random initial conditions are calculated numerically, and the probability of their contact with a collector is estimated. The results of Monte-Carlo simulations illustrate that larval self-propulsion may increase the probability of their contact with a collector by orders of magnitude compared to passive particles.

## Introduction

A number, which is called the *probability of settlement*, is often used in the biological literature to characterize the proportion of larvae that have settled on a specific substratum to the total number of larvae that could possibly settle on this substratum under the same biological and physical conditions [1].

It is common to distinguish between larval collectors of infinite extent (e.g., plane substrates) and finite size collectors. Our work examines settlement of larvae on a protruding isolated collector, a problem that, to date, has received much less attention despite its importance.

The probability of settlement is a product of the probability of contact with and the probability of attachment to a collector. Contact precedes attachment. Hence, the probability of contact represents the upper bound of the settlement probability. This number is also of a great interest in marine biology, especially if settlement follows the first contact event [10].

The geometry of natural finite size collectors is difficult to describe precisely. For achieving a better understanding of the complex settlement phenomenon it is also common to study settlement of larvae on/in bodies of well-defined geometry, such as tubes [2-3], plates [1], [4-6] or cylinders [7-9]. Our work examines contact of larvae with a long vertical cylinder located in an unbounded sea current.

Given that far from a collector the velocity of the sea current is typically much higher than the swimming velocity of a larva, the larva's motion toward a collector is generally regarded as transport of a passive particle. If a larva always behaves as a passive particle, then its contact with a collector can be studied within the framework of the theory of aerosols and hydrosols, in which the probability of contact of a large number of identical passive particles with a collector is termed the collector's *collision efficiency* [11]. A definition of the collision efficiency of passive particles which is adopted from the theory of aerosols and hydrosols [11]. For an unbounded flow it is defined as the ratio of the number of particles $n_c$ striking the collector to the number of particles $N$ that would have passed through it if they moved along ballistic trajectories (i.e., along straight lines) [11]:

$$E = \lim_{N \to \infty} \frac{n_c}{N} . \tag{1}$$

An important characteristic of a collector of particles is the rate of contacts of particles with the collector, i.e., the number of larvae that contact a characteristic area of a collector in a unit of time $t$:

$$\frac{dE}{dt} = CEU_{\infty}S , \tag{2}$$

where the concentration number $C$ is the number of larvae in a unit volume of the bulk water. The collision efficiency of a collector and the contact rate of collision are strongly related characteristics: (2) follows from (1). However, for biological applications the rate of contacts may be a more convenient characteristics than the collision efficiency of a collector which may be very low because tiny particles in the velocity field of a large collector move approximately along stream lines which do not cross the surface of the collector.

Due to the very low collision efficiency of collectors of tiny passive particles [11-14], some researchers consider water flow as a *"hydrodynamic impediment"* to larval settlement [15] (see also the corresponding argumentation in Fig. 1). Larva may increase its odds to make contact with a collector by increasing the radius of contact $\Delta$, e.g., by increasing the volume of its body or by protruding long mucous threads [10], [15]. A larva may also sense a collector and navigate toward it via attraction by various cues, including diluted chemical material or biofilms [16-19]. In turn, it can be argued that many larvae of marine invertebrates (e.g. the bryozoan *Bugula neritina*) are not capable of significantly increasing their contact radius, or may settle on clean surfaces [2-3], which seemingly excludes the role of chemical cues during the process of selecting a substrate for settlement.

In this work, we present a mathematical model of larval contact with a collector that differs from a model of passive contact by one factor. As in the concept of passive contact, we assume that the collector does not induce any cues, a larva does not know where the collector is located, and does not increase its contact radius. However, a larva is self-propelled. Although the words “larva” and “particle” are not synonyms, we use them interchangeably. We consider a larva as a self-propelled particle. In our formulation, the larva moves as an underwater vehicle equipped with a propulsor but whose navigation device is not functioning. If such a vehicle meets a large obstacle located in unbounded sea the sea current, e.g. in a large vortex, is the collision with the obstacle unavoidable or avoidable? If the same vehicle repeats its motion many times but with different initial conditions, what is the probability of collision? Similar questions are frequently asked in the theory of marine vessel control [20-21]; whereas similar aspects of the motion of a self-propelled marine larva in a large vortex were studied in [22-24]. Despite a huge difference in the scales of a marine vehicle and a larva, the vehicle and the larva moving in a fluid obey certain general laws of mechanics and hydrodynamics that have much in common for such different objects.

The structure of the paper is as follows. We first describe trajectories of *B. neritina* larvae measured in a laboratory flow tank. The analysis of the trajectories allowed us to develop a simple description of a larva's typical helical motion and estimate the relevant parameters of the helix, which are included in the mathematical model of motion of a larva. For computing the velocity field in the proximity of a collector, we used a commercial computational fluid dynamic (CFD) package, ANSYS-FLUENT 13.0 (hereafter AF13) [25], and the methods of classical boundary layer theory [26]. The probability of contact of larvae with a collector was calculated by using the Monte-Carlo method [27], i.e., by simulating a large number of larvae trajectories with random initial conditions. Next, we show that self-propulsion may greatly increase the odds of a larva to make contact even if the location of the collector is unknown.

# Results

Due to a huge variety of larval forms, a mathematical model of a larva's motion cannot be rather general. In each particular case, the parameters that may be necessary for mathematical modeling of larval motion can only be found experimentally. As a basis of our mathematical we chosed the bryozoan larva of *B. neritina*. If a larva can be considered as a mechanical object, we expect that the essential features of a mathematical model of motion of *B. neritina* may have much in common with mathematical models of the motion of other self-propelled larva.

## Parameters of a larva and of its typical trajectory

A photograph of the body of a *B. neritina* larva, as seen under an electron microscope, is shown in Fig. 2. The shape of the larvae is similar to a prolate spheroid with a length to maximal width ratio of 1.1 approximately. Such spheroid can be approximated by a sphere of the same volume i.e., an equivalent sphere with a radius of $d_p = 200 - 350\mu\text{m}$ [28-29]. Once the approximate volume of an equivalent sphere is known, the difference of the mean density of the body of a larva $\rho_p$ and the water density $\rho_f$ can be estimated using the sinking velocity of an immobilized larva, of the order of $V_t \approx 1\text{mm/s}$ [30]. Although the drag of a smooth sphere and of a ciliated larva are not the same [31] using Stokes' formula provides a correct order of magnitude of the ratio $\rho_p / \rho_f \sim 1.1$.

Trajectories of larvae in a laboratory flume are shown in Fig. 3. Enlarged sections of the trajectories are separately plotted in Fig. 4. In Figures 3-4 larvae typically moved along a path, which can approximately be described by a helix (Fig. 5).

A helix is drawn by the end of a line of length $a$, which rotates about an axis $o\xi$ with constant angular frequency $\gamma$ and translates simultaneously along this axis with constant velocity $V_S$. The angular frequency $\gamma$ and the temporal period of a helix $T$ are related as $\gamma = 2\pi / T$. When a certain point of a helix makes a full turn, it translates along the axis of the helix at distance $P$, which is called the pitch of the helix. The trajectory of a point of a helix can be described in a Cartesian coordinate system $o\xi\eta\zeta$ with unit vectors of the axes $(\mathbf{i}_\xi, \mathbf{i}_\eta, \mathbf{i}_\zeta)$ by a parametric dependence $\xi = V_S t$, $\eta = a\sin(\gamma t + \beta)$, $\zeta = a\cos(\gamma t + \beta)$ where $t$ is the time and $\beta$ is a phase angle. The velocity components of a point of a helix are given by the temporal derivatives of the coordinates of the point:

$$\mathbf{V}_h = V_S \mathbf{i}_\xi + a\gamma\cos(\gamma t + \beta)\mathbf{i}_\eta - a\gamma\sin(\gamma t + \beta)\mathbf{i}_\zeta \,. \quad (3)$$

In view of (3) the term "a larva's velocity" may refer to the two related but different velocities. One of them pertains to the mean velocity of a larva $V_S$ in the preferable direction of motion $O\xi$, termed here "the swimming velocity". The absolute value of the instantaneous velocity of the larva's motion along the helical path $V_h$ and the swimming velocity $V_S$ relate as:

$$V_h = \sqrt{V_S + (a\gamma)^2} \,. \quad (4)$$

It should be noted that a larva never moves along a regular helix with a straight axis $o\xi$. However, if the radius of curvature of the mean trajectory of a larva is much larger than the local pitch of the corresponding helix, then equations (3) can still be used as an approximate model of larval helical motion.

Two dimensional (2D) trajectories of a larva shown in Fig. 3-4 can be used for approximate estimating of the diameter of a helix $d_h = 2a$ and its temporal period $T$. Even if the actual direction of the swimming velocity of a larva is unknown, the projection of a cylindrical helix on the plane of a lens is confined by two parallel lines with the distance between them equal to the diameter of the helix (Figures 4-5). The temporary period of the helix can be estimated by calculating the time at which a larva moving along the helix makes a full turn. For example, using data from Figure 4, the radius of the helix $a$ can be estimated as approximately three diameters of a larva; whereas the temporal period of the helix $T$ can be estimated as approximately 1 s.

Estimating the swimming velocity $V_S$ from 2D images remains problematic because the mean trajectory of a larva is not in the plane of the lens (see Fig. 5). However, if the maximal velocity $V_h$, the period of a helix $T$ and the radius of a helix $a$ are known, then in view of (4) the swimming velocity $V_S$ can be calculated as:

$$V_S = \sqrt{V_h^2 - (a\gamma)^2} \,. \tag{5}$$

Wendt [28] suggested measuring $V_h$ by filming the motion of larvae in a shallow depth of field and in such way that only a small portion of a larva's trajectory is in focus. In such a case the distance which a larva covers in focus divided by the corresponding time gives an estimate of $V_h$. According to Wendt [28] for *B. neritina* $V_h \approx$ 5 mm/s. It such a case he swimming velocity $V_S$ can be estimated as approximately $V_S = 3$ mm/s. Qian et al. [2] estimate the swimming velocity of *B. neritina* as 1 mm/s; whereas we observed values $V_S$ varying from 0.5 to 6 mm/s.

It should be noted that the helical motion is a rough model of a real trajectory of a larva. Moreover, in many occasions larvae makes an unpredictable sharp turn, which can not be modeled rigorously. However, in a flow, the resulting velocity vector of a larva can be represented approximately as a vector sum of the velocity of the fluid and the velocity of a larva in still water. Because the flow velocity is higher than the swimming velocity of a larva then the larva's trajectory in a flow straighten (see Fig. 6 for more detail explanations). Thus, we assume that unpredictable irregularities of a larva's trajectories in still water do not influence significantly the general kinematical pattern of its motion in flow, the case that we study here.

## Mathematical model of the probability of contact

The problem of calculating the collision efficiency is related directly to the hydrodynamic problem of larval transport in the velocity field of a collector. In this context, two approaches can be used to calculate particle transport in flow: Eulerean and Lagrangian [32]. In the Eulerean approach, the particles are regarded as a continuous phase, which obeys the same laws as the carrier fluid. It is not trivial to include the hydrodynamic forces acting on a particle in the Eulerian approach.

In the Lagrangian approach, particles are treated as a discrete phase, in which each individual element moves in a carrier fluid under the action of hydrodynamic forces. The principal steps of the Lagrangian approach are as follows. First, the velocity field around a collector is computed. Then, a number of particles are ejected into the fluid.

The Lagrangian approach is computationally intensive and time consuming. However, it allows one to take into account all relevant hydrodynamic forces and to consider, in detail, the mechanics and hydrodynamics of contact. In the present work, we used the Lagrangian approach.

For laminar flows, the velocity field induced by a collector can be calculated using the Navier-Stokes equations. The presence in the flow turbulent pulsations of the imbedded sea turbulence makes it impossible to calculate the collision efficiency of a collector on a purely theoretical basis without invoking, directly or indirectly, experimental data. Because of these difficulties, we tacitly postpone the discussion on the influence of turbulence on the collision efficiency of a collector and first formulate the problem for laminar flow conditions. The effect of turbulence of the collision efficiency is discussed later.

## Equations of motion of a self-propelled particle

A collector of larvae is assumed to be a long vertical cylinder of diameter $D_c$ (radius $R$). The cylinder is located in unbounded sea current whose mean velocity vector $\mathbf{U}_\infty$ is normal to the cylinder's axis. The fluid velocity field in the presence of the cylinder is denoted as $\mathbf{U}$. It is assumed that a particle is small compared to the typical length scale of the variation of the velocity field induced by the cylinder. We considered cylinders with Reynolds numbers $\mathrm{Re}_c = U_\infty D_c / \nu$ such that $10^2 < \mathrm{Re}_c < 10^4$. In this case the flow at the front portion of the cylinder is laminar albeit, as it was noted above, it can be disturbed by embedded turbulence.

We consider a larva whose sinking velocity is much lower than the velocity of the sea current $U_\infty$. Under such circumstances, the distance, which a larva travels in time unit in a horizontal direction, is much larger than the corresponding distance of its sinking. Therefore, in order to simplify the problem formulation, we assumed that a larva moves in a horizontal plane.

The collision efficiency of the collector depends on the contact radius of a larva, $\Delta$. Larger radii of contact result in larger probabilities of contact. In our numerical simulations, we assumed $\Delta = d_p / 2$.

The trajectory of a larva was described in the coordinate systems shown in Fig. 7. Following the common practice adopted in the theory of aerosols, to calculate the collision efficiency of a cylinder, the initial longitudinal coordinates of all particles $X_o(0) = X^0$ may be chosen to be equal. It is also implied that at a distance $X^0 \gg D_c$, the disturbances introduced by the cylinder in the flow are small compared to the velocity of the undisturbed flow $U_\infty$.

For passive particles, the initial transverse coordinates $Y^0$ may be chosen to be random numbers that are uniformly distributed in the segment $(-D_c/2, D_c/2)$; whereas the initial angle $\varphi(0) = \varphi^0$ between the axes $ox$ and $OX$ does not influence its trajectory. For a living larva, the angle $\varphi^0$ determines the initial direction of the velocity $\mathbf{V}(0)$ (Fig. 8).

It is assumed here that far from the collector, larvae move in a certain preferable direction within an angle $-\Phi < \varphi^0 < \Phi$ ($0 \le \Phi \le \pi$) where the initial angles $\varphi^0$ are random numbers distributed within the interval $(-\Phi, \Phi)$ uniformly. In such a formulation, the angle of the preferable direction of motion becomes one of the problem parameters whose bearing on the collision efficiency becomes a part of the investigation.

There are three main mechanisms of the collision of a passive particle with a collector: inertial impaction, direct interception and Brownian diffusion. The inertial collision is determined by the inertial parameter representing the product of the Stokes number

$$\mathrm{Stk} = \frac{\rho_p}{\rho_f} \frac{d_p^2 U_\infty}{18 \nu D_c} \quad (6)$$

and a factor $(1 + \rho_f / 2\rho_p)$, where $\nu$ is the water kinematic viscosity [33]. Typically, for many small marine larvae (including *B. neritina*), the inertial Stokes parameter is much less than 1/8, a threshold value below which contact of an inertial point with a cylinder does not take place [11]. For the problem parameters which are adopted here the Stokes number is of order of $\sim 10^{-2} - 10^{-3}$.

Diffusional deposition results from the random Brownian movement of water molecules that bombard particles and may bring them to a collector. Given that the own velocity of *B. neritina* is many orders of magnitude higher than its velocity, which might be induced by the Brownian motion, the mechanism of diffusional deposition is neglected in the proposed mathematical model of contact.

The trajectory of a particle can be described by a derivative of the radius-vector $\mathbf{r}$ with respect to time $t$ (see Fig. 7):

$$\frac{d\mathbf{r}}{dt} = \mathbf{V}, \quad (7)$$

where the particle's velocity vector $\mathbf{V}$ can be calculated by solving its differential equations of motion.

It is assumed that the motion of a particle with respect to the fluid is slow and that the Reynolds number $\mathrm{Re}_p = |\mathbf{V} - \mathbf{U}| d_p / \nu$ is low. Within the framework of low Reynolds number hydrodynamics and by neglecting the Basset integral, the equation of motion of a particle in a simplified form can be written as follows [33]:

$$m_p \frac{d\mathbf{V}}{dt} = 3\pi\mu d_p (\mathbf{U} - \mathbf{V}) + \frac{1}{2} m_f \frac{d}{dt}(\mathbf{U} - \mathbf{V}) + m_f \frac{d\mathbf{U}}{dt} + \mathbf{g}(m_P - m_f) + \mathbf{F}_T. \quad (8)$$

Here, $m_p$ is the mass of the particle, $m_f$ is the mass of the fluid in the volume displaced by a sphere , $\mu$ is the water viscosity and $\mathbf{g}$ is the vector of the acceleration of gravity.

The last term in the right hand side of (8) represents the thrust of the particle. The relationship between the velocity $\mathbf{V}_h$ of a particle with respect to water and the thrust $\mathbf{F}_T$ can be expressed by:

$$\mathbf{F}_T = 3\pi\mu d_p k_p \mathbf{V}_h \quad (9)$$

where a correction factor $k_p$ takes into account the non-sphericity of a larva's body and its cilia. Given that even for an immobilized larva, calculating the drag of its ciliated body may be problematic [31], in order to proceed and to obtain at least an approximate estimate of the drag of a larva's body, it is plausible to replace it with an equivalent sphere of diameter $d_p$. For a smooth sphere the correction factor can be expressed as a function of the Reynolds number $\mathrm{Re}_h = |V_h - U| d_p / \nu$ as follows: $k_p = (1 + 0.158\,\mathrm{Re}_h^{0.667})\ (\mathrm{Re}_h < 500)$ [33]. For *B. neritina,* $\mathrm{Re}_h$ is on the order of one, and the correction factor is of the same order of magnitude.

The thruster of a larva creates not only the force $\mathbf{F}_T$ but also a torque hydrodynamic moment $\mathbf{M}_p$. Due to this moment, a larva rotates in still water with a certain angular velocity $\boldsymbol{\omega}_p$. For a smooth sphere rotating with low Reynolds numbers $\mathrm{Re}_\omega = |\omega_p| d_p^2 / \nu$, this moment can be calculated as $\mathbf{M}_p = \pi d_p^3 \mu \boldsymbol{\omega}_p$ [35].

The water flow in the area influenced by the collector is subject to shear. A small particle placed in a viscous shear flow is subject to a hydrodynamic moment $\mathbf{M}_s = \pi d_p^3 \mu \boldsymbol{\omega}_s$, which causes the particle to rotate with an angular velocity $\boldsymbol{\omega}_s$ [34]. The latter is equal to approximately half the vorticity vector $\mathrm{rot}\,\mathbf{U} = 2^{-1}(\partial U_X / \partial Y - \partial U_Y / \partial X)$, i.e., $\boldsymbol{\omega}_s = 2^{-1} k_\omega \mathrm{rot}\,\mathbf{U}$, where for a low rotational Reynolds number $\mathrm{Re}_\omega$, the correction factor $k_\omega$ mainly depends on the translational Reynolds number $\mathrm{Re}_p$ and for $\mathrm{Re}_p \le 200$ is of the order of one [36]. Under the combined action of the two torque moments, $M_p$ and $M_s$, a particle begins to rotate with an angular velocity $\boldsymbol{\omega}$, which in turn is associated with a hydrodynamic moment of fluid resistance to this rotation $\mathbf{M}_f = -\pi d_p^3 \mu \boldsymbol{\omega}$.

The equation of the conservation of the angular momentum of a rotating particle is:

$$J \frac{d\boldsymbol{\omega}}{dt} = \mathbf{M}_f + \mathbf{M}_p + \mathbf{M}_s, \tag{10}$$

where $J = \pi \rho_p d_p^5 / 60$ is the mass moment of the inertia of a sphere. Substituting the expression for $\mathbf{M}_f$, $\mathbf{M}_p$ and $\mathbf{M}_s$ into (10), we obtain the differential equation of a rotating particle with respect to $\boldsymbol{\omega}$:

$$\frac{\rho_p d_p^2}{60\mu} \frac{d\boldsymbol{\omega}}{dt} + \boldsymbol{\omega} = \boldsymbol{\omega}_f + \boldsymbol{\omega}_p. \tag{11}$$

The rotation of a particle with angular velocity $\boldsymbol{\omega}_p$ due to its helical motion is considered here as a prescribed one. Once the vector of the swimming velocity $\mathbf{V}_S$ and the parameters of the helix are known, then, in view of (3), the right hand side of (8) is defined. In such a case, the influence of the angular velocity $\boldsymbol{\omega}_p$ on the velocity of the particle $\mathbf{V}$ can be neglected in the first approximation.

Equations (8) and (11) can be further simplified. Different orders of magnitude of the constituting terms can be revealed by introducing dimensionless variables $\mathbf{W}^+ = (\mathbf{V} - \mathbf{U}) / U_\infty$, $\mathbf{U}^+ = \mathbf{U} / U_\infty$, $\mathbf{V}_h^+ = \mathbf{V}_h / U_\infty$, $t^+ = U_\infty t / D_c$, $\mathrm{Fr} = U_\infty^2 / g D_c$, and $\delta = 1 - \rho_f / \rho_p$, yielding:

$$(1+\frac{\rho_{\rm f}}{2\rho_p}){\rm Stk}\frac{d\mathbf{W}^+}{dt^+}+\mathbf{W}^+=-\delta{\rm Stk}\frac{d\mathbf{U}^+}{dt^+}+\frac{\delta{\rm Stk}}{{\rm Fr}^2}\frac{\mathbf{g}}{g}+\mathbf{V}_h^+, \qquad (12)$$

$$\frac{3}{10}{\rm Stk}\frac{d\boldsymbol{\omega}}{dt^+}=(\boldsymbol{\omega}_f-\boldsymbol{\omega}), \qquad (13)$$

In (12)-(13) the Stokes number Stk and the buoyancy parameter $\delta$ are small quantities of the first order of magnitude. By neglecting in (12) their product, i.e. small terms of the second order of magnitude, we obtain the following equation of motion of a tiny self-propelled particle:

$${\rm Stk}\frac{d\mathbf{W}^+}{dt}+\mathbf{W}^+=\mathbf{V}_h^+, \qquad (14)$$

or

$$\tau\frac{d\mathbf{V}}{dt}+\mathbf{V}=\mathbf{U}+\mathbf{V}_h, \qquad (15)$$

where $\tau=(1+\rho_f/2\rho_p)\rho_p d_p^2/18\mu$ is the relaxation time (of the order of $10^{-2}{\rm s}$ for the problem parameters). By neglecting all small terms in (12)-(13) and returning to dimensional notations, we receive purely kinematic equations of motion:

$$\mathbf{V}=\mathbf{U}+\mathbf{V}_h, \qquad (16)$$

and

$$\frac{d\mathbf{r}}{dt}=\mathbf{V}, \quad \frac{d\varphi}{dt}=\omega_f. \qquad (17)$$

Eq. (16)-(17) represent the simplest model of the motion of a self-propelled larva with negligible inertia. The helical motion of the larva is incorporated into (16) through the velocity $V_h$, given by (3). Further, depending on a purpose, we use equations of motion of an inertial larva or a larva which is massless.

## Efficiency of a collector

Because the motion is considered in a horizontal plane and $\rho_p/\rho_f\sim 1$ within the framework of our mathematical model the probability of contact $E$ is a function of four non-dimensional parameters: $V_S^+=V_S/U_\infty$, $d_p^+=d_p/D_c$, $\Phi$ and ${\rm Re}_c$. The relative importance of each of them on the probability of contact can be established by numerical simulations.

In [12] the collision efficiency of a cylindrical collector of passive particles for $0.1<{\rm Re}_c<50$ was analyzed using CFD package COMSOL. In our work we analyzed the collision efficiency of a cylindrical collector of passive particles using CFD package AF13. To analyze the probability of contact of self-propelled particles with a cylinder for $100<{\rm Re}_c<500$ we used the theory of the boundary layer [26].

AF13 CFD package enables the calculation of both the velocity field around a cylinder and the particles' trajectories where the motion of an individual passive spherical particle is described by a simplified equation (15):

$$\tau \frac{d\mathbf{V}}{dt} + \mathbf{V} = \mathbf{U}. \tag{18}$$

Fig. 9 illustrates the streamlines of the flow and the trajectories of particles in the vicinity of a cylinder which are almost indistinguishable within the thickness of the plotted line. It is an indication that for the problem parameters adopted here the inertia of a particle is negligible, and contact may be considered as a result of the direct interception solely.

We did not attempt to present systematic data of the probability of contact for passive particles and all of the various problem parameters. This computational task is far beyond the scope of the present paper. As stated in the "Introduction", our aim was to evaluate the role of larval self-propulsion on a collector's collision efficiency. For this purpose, we used the approximate methods of the boundary layer theory instead of the direct solution of the Navier-Stokes equations.

In particular, for calculating the velocity field on the front part of a cylinder, both the Blasius and the Pohlhausen's method provide accurate results sufficient for intended purposes, except for the nearest vicinity of the flow separation point [26]. Given that most collisions of passive particles occur at the upward portion of a cylinder [11-14], we considered this inaccuracy as minor. Here, we used Pohlhausen's method because it requires less computation compared to the Blasius method [26]. Outside of the boundary layer, the fluid velocity field around a cylinder is calculated as for an inviscid irrotational flow. Because the results of CFD computations illustrated a negligible effect of the particles' inertia on their trajectories, in most computations, we used equations (16)-(17) although for justification purposes part of numerical simulations was carried out using (13)-(14).

Examples of the calculated trajectories of larvae are shown in Fig. 10. To estimate the probability of contact of larvae with a collector, it is necessary to calculate a large number of trajectories for $N_p$ particles. The influence of this number of the probability of contact is shown in Fig. 11**.** The collision efficiency calculated for different angles of the preferable direction of motion $0 \le \Phi < \pi$ is plotted in Fig. 12.

The above numerical examples give clear evidence that the ratio of a larva's swimming velocity to the flow velocity $V_S / U_\infty$ strongly influences the collision efficiency of a collector. However, these results were obtained for constant values of the ambient flow velocity and variable swimming velocity, whereas the same ratio $V_S / U_\infty$ can be obtained for constant $V_S$ and variable $U_\infty$. The question arises of whether, in both cases, the collision efficiency coefficient remains the same at least approximately. Fig. 13 provides a positive answer to this question in the range of the flow velocities $U_\infty = 1\text{-}10\ \text{cm/s}$ and a larva's swimming velocities $V_S = 0.05 - 0.5\ \text{cm/s}$.

## Collision efficiency of a collector with embedded sea turbulence

There are very little references in the literature to the collision efficiency of cylinder under turbulent flow conditions. In this section we are attempting to estimate the collision efficiency of a cylinder under turbulent flow conditions by using CFD methods. Most of them that are used to calculate the parameters of the turbulent flow, are based on the Reynolds average Navier-Stokes equations (RANSE) that are combined with equations of the kinetic energy $k$ and the energy dissipation rate $\varepsilon$ (e.g., $k-\varepsilon, k-\omega$ and Reynolds Stress models).

A frequently used model to describe particle motion in a turbulent flow is the eddy interaction model (EIM) developed in [37], which is represented in most of the AF13 codes (see also [38]). According to [37], the integral scale $L_\varepsilon$ of the largest eddies feeding the turbulent system far from the collector, the turbulent kinetic energy $k$ and the energy dissipation rate $\varepsilon$ are related as $L_\varepsilon = C^{3/4}k^{3/2}/\varepsilon$, where $C = 0.09$ for $k$-$\varepsilon$ and $k$-$\omega$ models. Far from the collector, the turbulent intensity $I$ of the embedded sea turbulence and the mean fluid velocity $\bar{U}_\infty$ are related as $k = 1.5(\bar{U}_\infty I)^2$. If the characteristic size of a larva $d_p$ is much less than the Kolmogorov length $\eta_k$ [39], then the larva sits in a laminar flow, and the hydrodynamic forces acting on it can be calculated as for a laminar flow. The ratio of the largest scale $L_\varepsilon$ of the turbulent flow to the smallest scale $\eta_k$ are related through the Taylor micro-scale Reynolds number $\mathrm{Re}_\lambda$ $L_\varepsilon/\eta_k = \left(\mathrm{Re}_\lambda^2/15\right)^{3/4}$. Fully turbulent flow requires $\mathrm{Re}_\lambda$ to be larger than approximately 100; the highest Reynolds numbers measured in tidal channels are $\mathrm{Re}_\lambda \approx 2000$ [40-42]. Table 2 presents typical values of the turbulence in the sea upper mixing layer [40-42].

Once the diameter of a particle is less than the Kolmogorov length and the rest of the parameters of turbulence are also defined, AF13 enables the calculation of the trajectories of a large number of particles that begin their motion in the nodes of the computational grid. In a turbulent stochastic random walk model from one node may start a number of particles. Due to the randomization of the problem, all their tracks are different.

We computed the collision efficiency of a cylinder with embedded turbulence for $\mathrm{Re}_c = 500$. It is known that even for a laminar flow with $\mathrm{Re}_c > 200$ the wake of a long cylinder is unsteady and can not be considered as two dimensional. In a flow with embedded turbulence in its wake are definitely three dimensional. Therefore our 2D calculations can not reproduce faithfully the three dimensional (3D) structures of the wake of the cylinder. However, because we consider collisions of particles with a vertical cylinder, it is expected that three dimensional structures will not affect significantly the collisions on the front part of the cylinder where most of them take place. The computed values of the probability of contact $E$ obtained using $k$-$\omega$ model of turbulence for realistic parameters of the sea turbulence are given in Table 2.

## Discussion

The general features of contact of a larva with a protruding collector are explained in further detail in schematic Figures 14 and 15. Motion of a larva in a vortical flow and its deviation from a streamline of a collector due to rotation is illustrated in Fig. 14. Fig. 15 illustrates the typical scales of the contact problem. Far from a collector, the trajectory of a larva is determined by the sea current and the swimming velocity of a larva. In area 1 the hydrodynamic disturbances $\mathbf{U}_c = \mathbf{U} - \mathbf{U}_\infty$, which are induced by a collector at a large distance $r$ from it, decay according to the dipole approximation as $U_c \propto U_\infty (R/r)^2$ [43]. In this area the influence of the fluid viscosity on the hydrodynamics is weak and the velocity field of the collector can be calculated as for a potential flow. The larva moves approximately along a corresponding streamline or deviates from it depending on the initial angle of the larva's of swimming velocity and on the ratio of its swimming velocity to the velocity of flow. In the boundary layer of the cylinder 2 of a thickness $\delta\,(d_p \ll \delta \ll D_c)$ the fluid viscosity is essential. In the boundary layer the larva translates, rotates due to the flow vorticity, and further deviates from the fluid stream line.

From the results of the presented numerical calculations, it follows that the collision efficiency of a collector is a growing function of the non-dimensional parameter $V_S / U_\infty$. Strictly speaking, this dependence is slightly nonlinear. However, it can be linearized and represented approximately as:

$$E \approx A + B\frac{V_S}{U_\infty}, \tag{19}$$

where the coefficients $A$ and $B$ are weakly dependent on the ratio $V_S / U_\infty$. Substituting (19) into (2) gives:

$$\frac{1}{CS}\frac{dE}{dt} \approx AU_\infty + BV_S\,. \tag{20}$$

The asymptotic formula (20) can be interpreted as follows: for a given unit of the concentration number, the amount of particles that are collected on a unit area of a collector is proportional to the velocity of the water flow far away from the collector and to the swimming velocity of a larva. Both components of (20) are of the same order of magnitude if $AU_\infty \propto BV_S$ or if $V_S / U_\infty \propto A/B$. Our numerical simulations demonstrated that for the problem parameters adopted here, $A \ll B$. Thus, we conclude that even a rather low larval swimming velocity may drastically influence the larval contact rate with a protruding collector.

Collision efficiency of cylindrical collectors of passive particles under laminar flow conditions for Stokes numbers less than 0.1 and $100 < \mathrm{Re}_c < 1700$ vary between $10^{-2} - 10^{-3}$ percents [44]. For a relatively small ratio of the swimming velocity of a larva to the flow velocity of order of 0.1 the probability of contact is of order of $\sim 2\%$ (Fig. 13, $E \sim 2\%$ for $V_S / U_\infty = 0.1$)). Thus, even weak swimming increases the probability of contact by 2-3 orders of magnitude.

These strong results pertain to laminar flow conditions; whereas real sea flows tend to be turbulent. Experimental measurements of the collision efficiency of cylinders under turbulent conditions for Stokes number less than 0.1 are extremely rare. The available experimental data, which may be relevant to our problem parameters, shows that for Stokes numbers of the problem less then 0.1 turbulence of intensity $I < 7.5\%$ has no effect on forward collision efficiency of cylinders [44], [45].

Estimates given in Table 2 illustrate that for the realistic problem parameters turbulence of intensity $I \sim 23\%$ may increases the collision efficiency of a cylinder also by orders of magnitude compared to the laminar flow conditions. However, for $V_S / U_\infty \sim 0.1$ the collision efficiency of a collector of passive particles is about 3 times less than the collision efficiency of the same collector of self-propelled particles (Fig. 13). Note that for our simulations we used strong sea turbulence with r.m.s of the velocities of the turbulent fluctuations of the order of 1 cm/s. This velocity is higher than the swimming velocity of the larva under consideration.

It should be stressed that the effect of sea turbulence on the collision efficiency of collectors of passive particles and marine larvae is still poorly understood, and more works are needed to clarify this important problem.

Our approach is based on observations of the trajectories of slow swimming and almost spherical bryozoan larva *B. neritina*. Thus, a question arises whether in the context of the contact problem this specific larva is a representative example for other species of marine larva. As long as a larva can be viewed as massless self-propelled particle, its motion can be described by kinematic equations similar to (16)-(17). It is not unlikely that they can be applied to other larva, which move with and rotates in a shear flow.

We propose a mechanistic and minimalist mathematical model of larval contact with a collector. It takes into account the details of the flow around the collector and the swimming of a larva in a preferable direction of motion. It does not include larval active behaviour as a reaction to light, to various dissolved chemical compounds around the collector, to turbulence as a possible trigger of larval swimming, and other various cues.

In this context it should be noted that two alternative approaches to the problem of larval contact with a collector are possible. The first is purely mechanical. Within the framework of this alternative, the rate of contact of larvae with a protruding collector can be calculated theoretically, though inevitably rather approximately. That is the approach that we have taken. However, it can be argued that a larva does not behave as a mechanistic object but simply swims deliberately across the direction of the flow gradient. This second behavioristic alternative, an empirical one, is not the subject of the present paper. However, in the mechanistic model of larval motion, which we have suggested here, at least one important behavioristic aspect is presented: a larva's self-propulsion.

We demonstrated that self-propulsion can greatly increase the larva's odds to make contact with a protruding collector, even if the self-propulsion is weak, and the location of the collector for the larva is unknown. From a larva's perspective, to reach a collector and to make contact with it, it is vitally important to be active.

## Materials and methods

### Collection of larvae

Sexually mature colonies of *B. neritina* were collected from floating docks in the Marina of Tel-Aviv, Israel, during the fall and winter period of 2009 and 2010 and again during the spring of 2010 following the methodology described in [3], [28]. Colonies were maintained in one liter plastic containers covered with aluminum foil to block light, which induces release of the larvae. Containers were transferred to darkened glass aquaria, where they remained in the dark with no supplemental food at approximately 25°C for a 24-h acclimatization period until the larvae were harvested. Larvae for each experiment were obtained from several colonies to foster genetically heterogeneous populations. Parent colonies used for experiments were maintained in the laboratory and harvested after 3-5 days. Larval release was induced with light. Colonies were removed from the dark, placed in 0.5 liter glass bowls with artificial seawater (approximately 39 ppm, prepared with Red Sea Coral Pro Salt) and then exposed to fluorescent light (two 28 W Semi Spiral Day Light lamps, 153 mA each). Larvae appeared within 10 min of illumination, and release was complete by 1 h. Because most larvae of *Bugula spp.* are positively phototactic on release, they aggregated at the illuminated side of dishes, a behavior that facilitated collection [28]. Larvae were collected using a glass pipette and placed in approximately 0.2 l before being transferred into the experimental flow tank.

## Recording the trajectories of larvae

The motion of *B. neritina* larvae was observed in a plane Couette flow apparatus, which consists of a transparent glass aquarium of 1.5x0.2x0.2 m and an electric motor rotating two vertical cylinders. The rotating cylinders move a closed flexible PVC belt. The aquarium is filled with a solution of rectified pure water and sea salt in a prescribed proportion. Larvae were introduced into the experimental tank using a glass pipette. Immediately after release, larvae started to move and did not stop unless they attached to the wall of the tank or to the belt. Larvae trajectories were recorded in still water and in a flow from above and from the side of the tank. Three different types of high-definition digital video cameras were used to film the moving larvae: a Mikrotron 1364 (500 frames/s), Canon 7D (60 frames/s) and Nikon 90 (24 frames/s), all equipped with macro lenses with 60 and 100 mm focal lengths. The filmed area was illuminated by disperse low light provided by four identical halogen lamps located approximately 1 m above the free surface of the water. The sensitivity of the cameras' sensors varied between 1600-6000 units of ISO equivalent. In most cases, the size of the photographed area was approximately $3 \times 4\,\text{cm}$, and the depth of field varied from a few millimeters to approximately 4 cm.

## Processing of trajectories of larvae in a flume

Trajectories of the larvae were digitized using the Image Processing Toolbox of Matlab (http://physics.georgetown.edu/matlab). On each frame, the software analyzed the centroid of the object detected by the subtraction method, and the path of a larva was formed as a consequence of centroids using the Matlab track function.

## Mathematical modeling

Analytic calculations were performed with pencil and paper. Numerical computations were performed by using commercial CFD package ANSYS-FLUENT 14, C and FORTRAN-77 programming languages.

The algorithm of the Polhasen's method [26] was programmed in Fortran 77 and executed under Windows using PCs. The differential equations of motion of a larva (15)-(17) were solved numerically using the method of Runge-Kutta of the 4-th order. The adaptive step of integration of the differential equations was at least 10 times less than $d_p / V_S$, the time which a larva needs to cover a distance equal to its characteristic length. Data fitting was performed in MatLab 7.

CFD numerical simulations were performed using a multiprocessor workstation SiliconGraphics CPU Intel Xeon with 16 GB Memory. Calculation of the collision efficiency of a collector of passive particles was carried out using AF13 with a grid with quadrilateral cells. Particles start their motion far from the cylinder, in the nodes of a computational grid with cells clustered to the cylinder. In the numerical examples provided here the maximal number of quadrilateral cells was 6,939,645 with the average element quality larger then 0.70. The average size of a quadrilateral cells adjusted to the cylinder was about $h = d_p / 200$. From each cell 20 particles begin their motion to the cylinder. The criteria of contact of a particle with the cylinder was formulated as $(l - d_p / 2) \le \Delta + h$, where $l$ is the distance from the center of the particle to the cylinder. The sensitivity of the results to the size of the mesh was justified by comparing the calculated drag coefficient of the cylinder $C_D$ with available experimental measurements.

# Figures

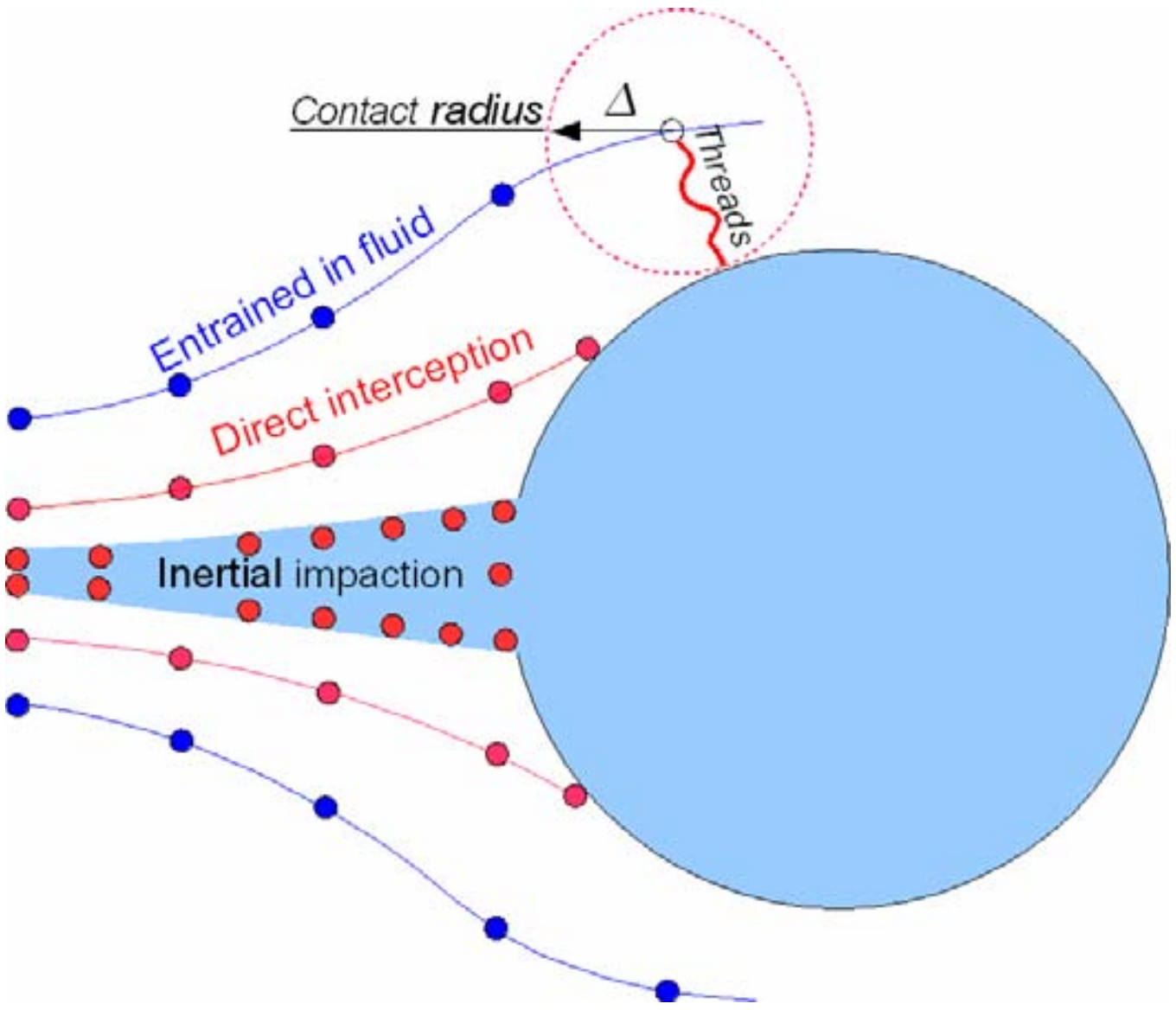

**Figure 1. Schematic model illustrating the contact of a tiny larva with a large collector.** The larva is "entrained" in the flow and follows the fluid streamline. The streamline does not cross the surface of the collector. Contact may take place only if the trajectory of a larva declines from a curvilinear streamline due to the larva's inertia (inertial impaction) or due to its finite size (direct interception). The inertia of a tiny, slow larva is small. For a very low ratio of the characteristic size of the larva to the characteristic size of the collector, the probability of the inertial impaction or direct interception is also very low. Given that not all larvae that contact the collector attach to it and that not all attached larva survive, the odds of larvae to colonize the collector are seemingly rather low.

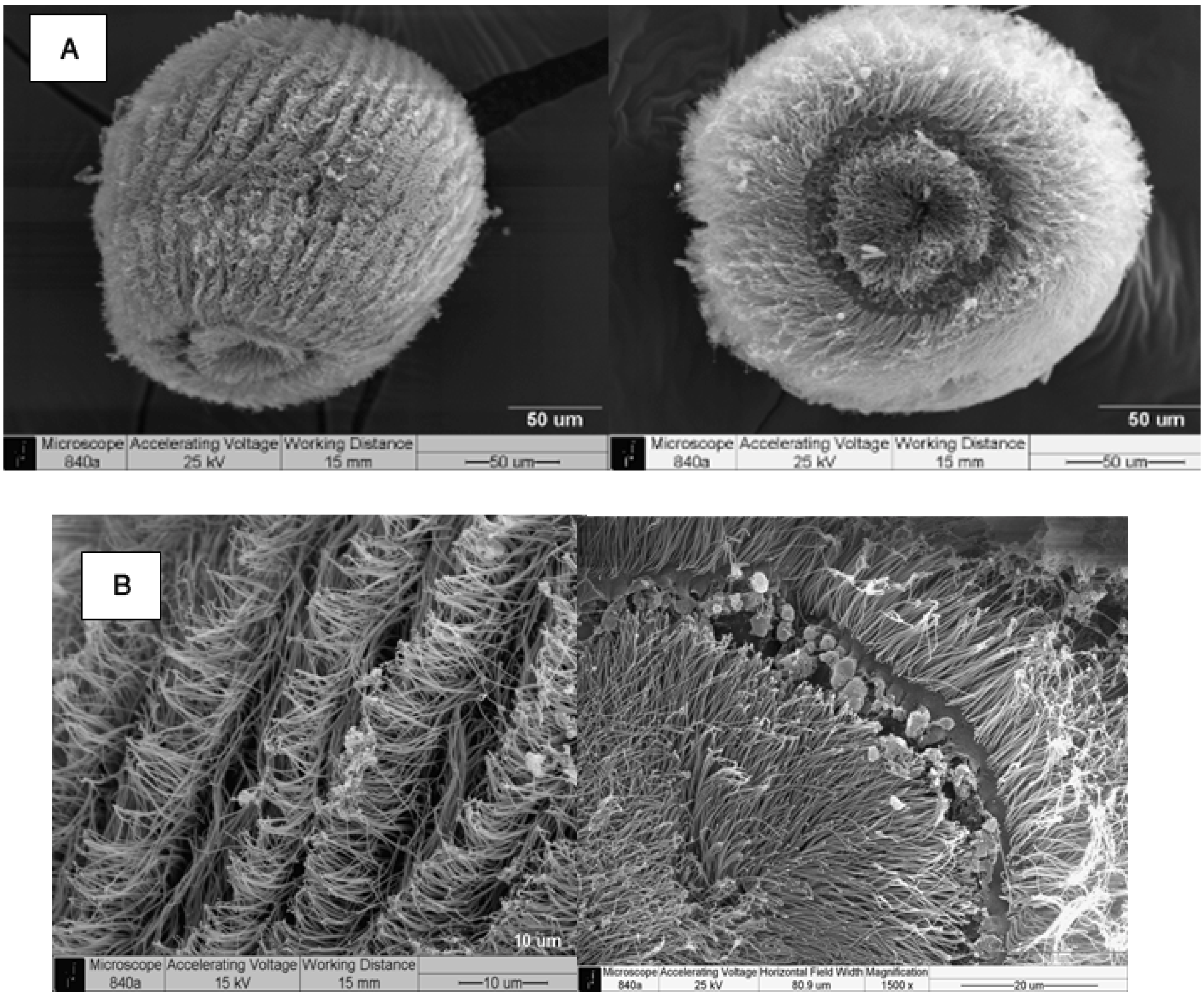

**Figure 2. Electron microscopy of *B. neritina* larva**: A. General view; B. Structure of the ciliated body surface with hair length estimated at 10 $\mu$m.

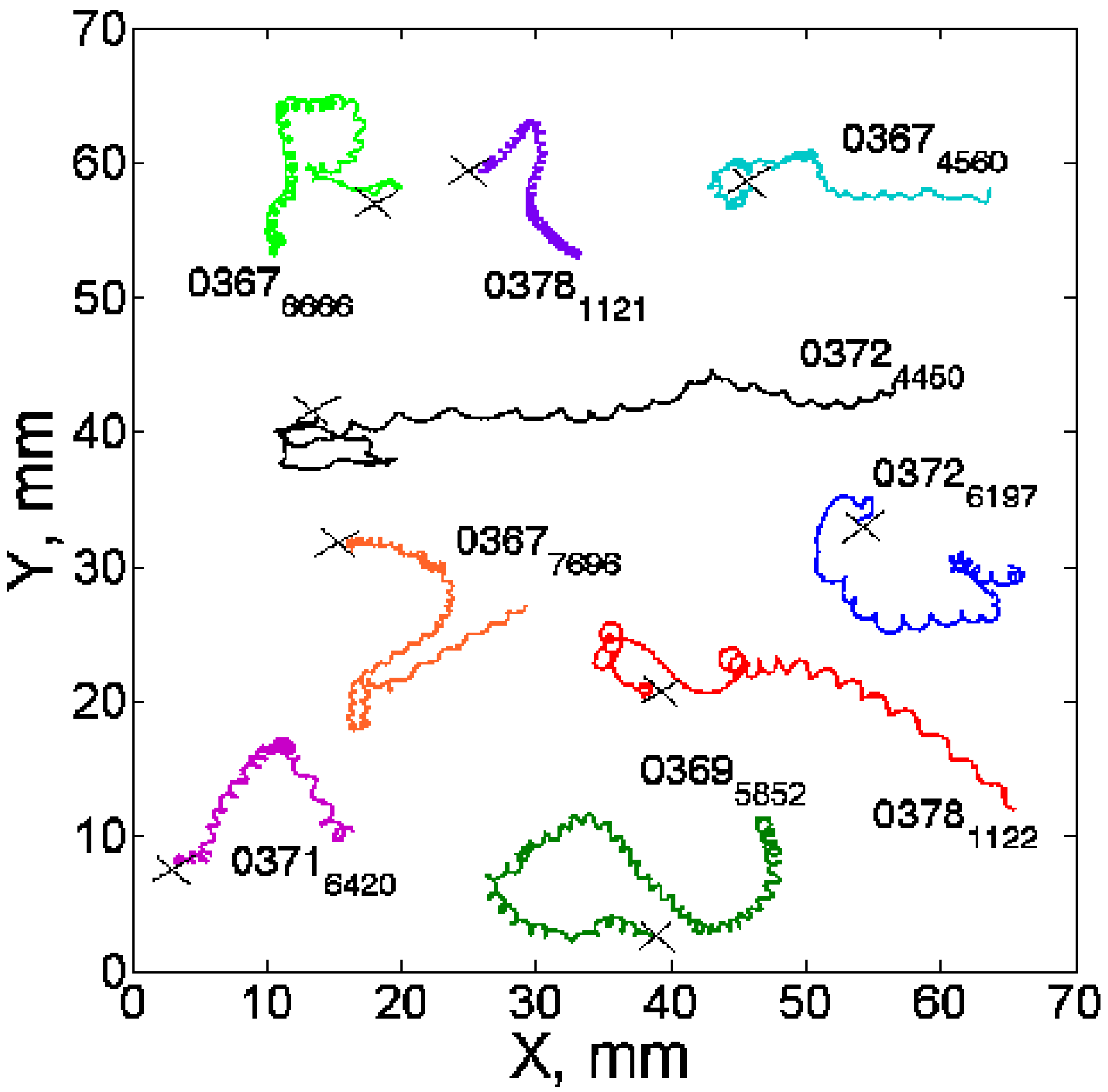

**Figure 3. Trajectories of larvae in still water.** Because of the additional illumination to one side of the tank larvae move in some preferable direction of motion (from left to right in the figure). × denotes the beginning of the trajectory.

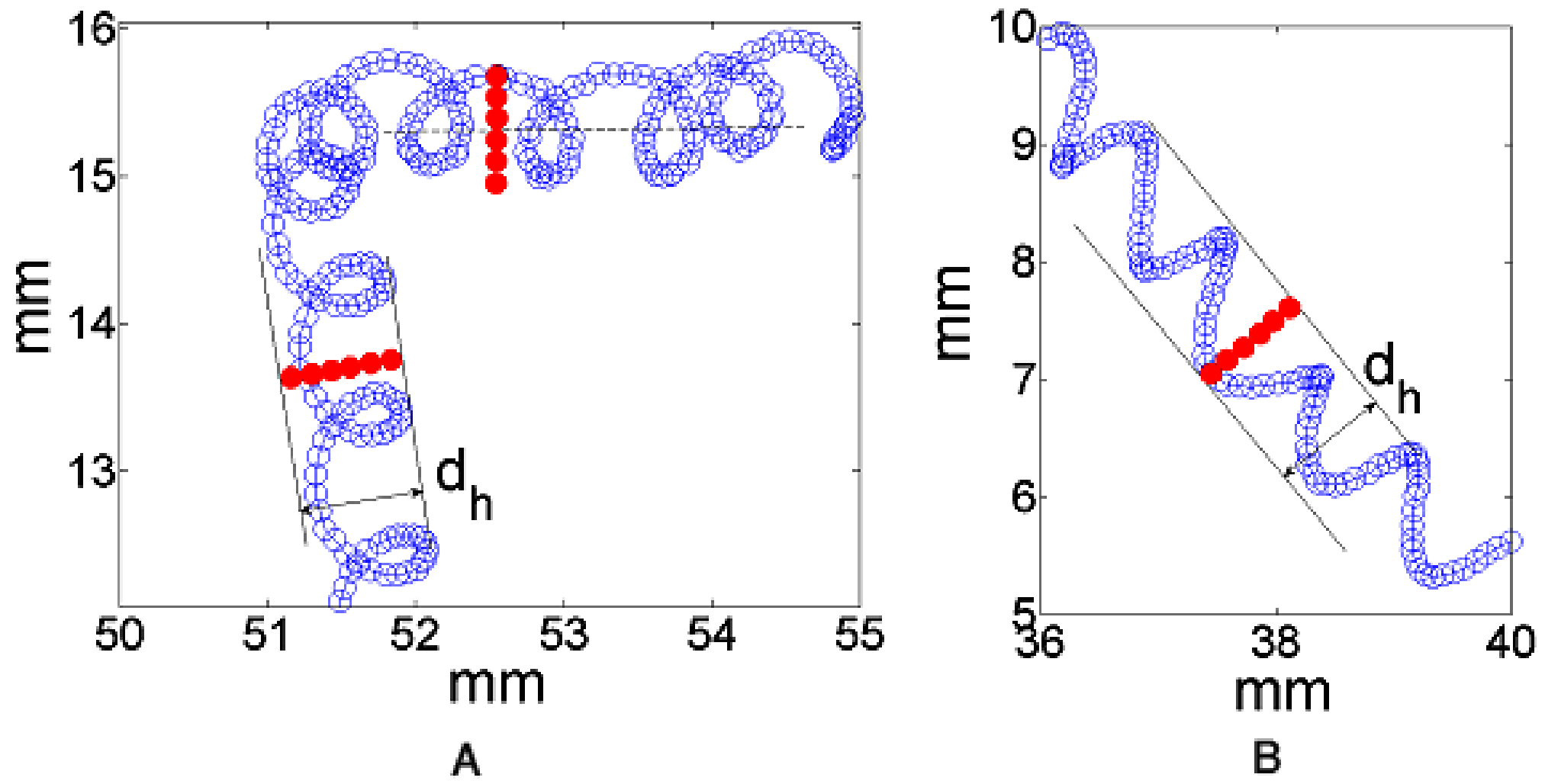

**Figure 4. Enlarged parts of trajectories:** A. $0367_{6666}$ ; B. $0369_{0582}$ . Open circles show the consequent position of a larva at a resolution of 1/24 s. Filled circles depict the estimated transverse displacement of larvae in the direction perpendicular to the preferable direction of swimming measured in diameters of a larva. The distance $d_h = 2a$ between two parallel lines bounding the trajectories can be considered as geometric characteristic of the helix.

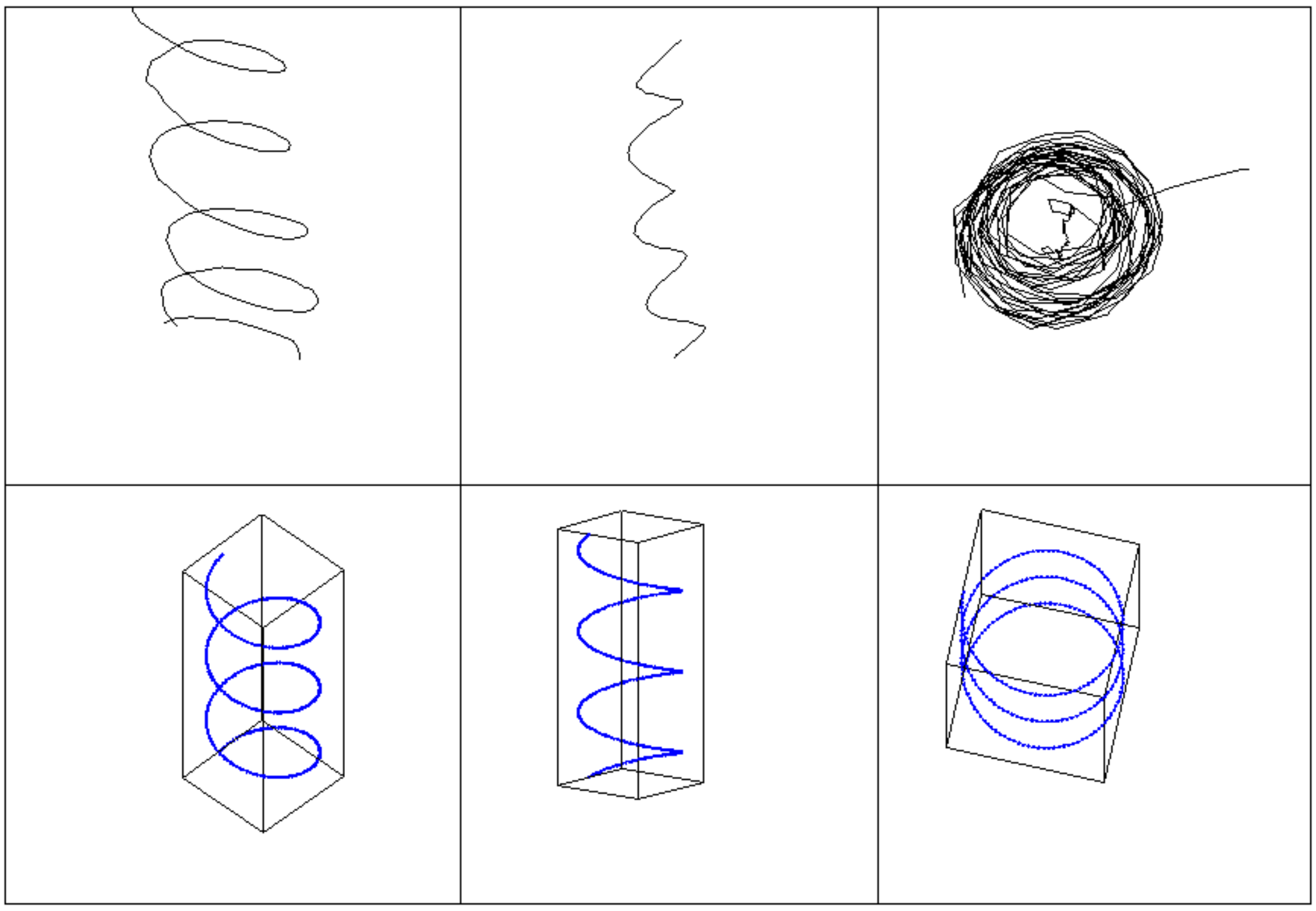

**Figure 5. Patterns of larval motion.** The upper row represents experimental 2D larval trajectories. In the lower row, the 3D helices are seen from different angles.

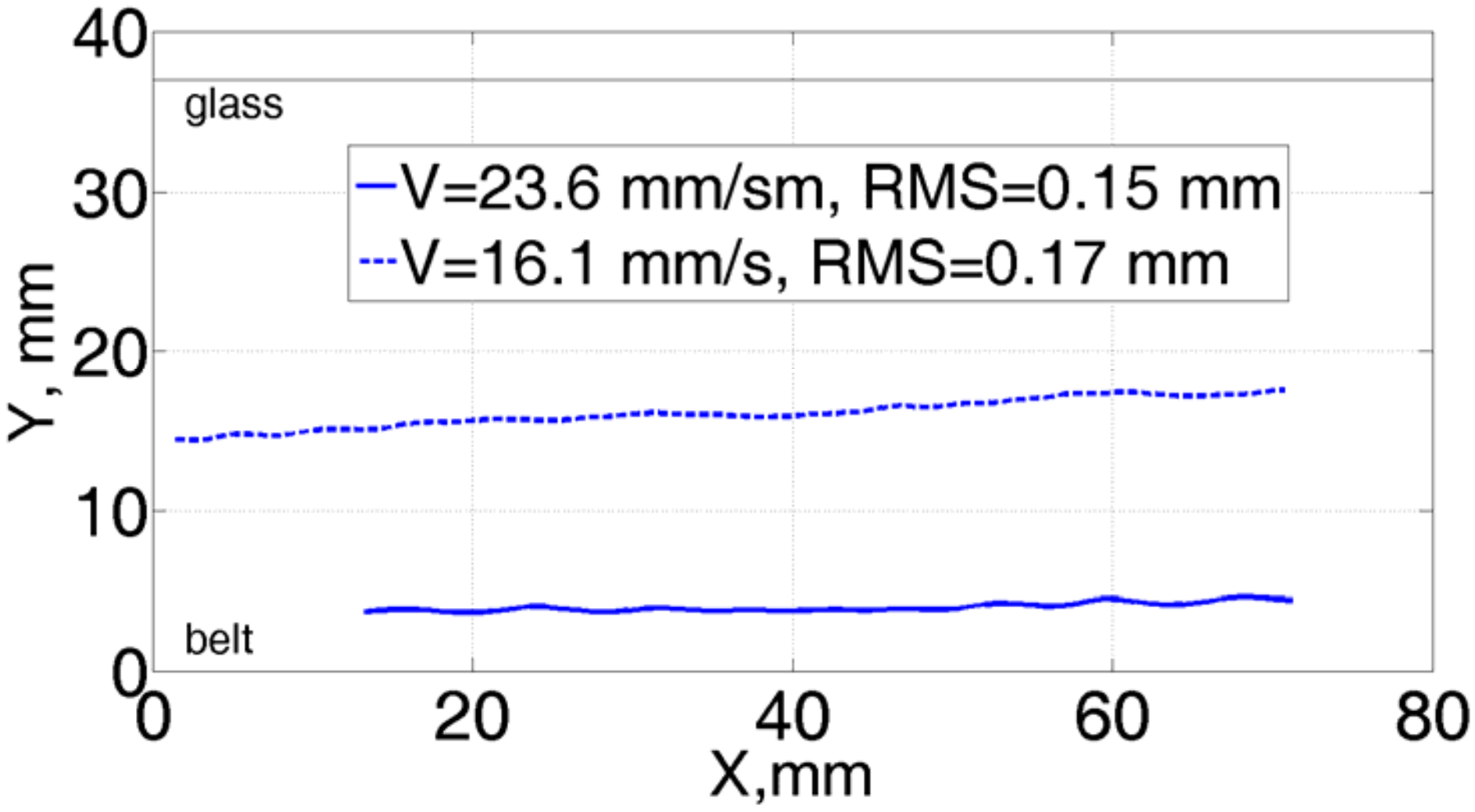

**Figure 6. Motion of larvae in Couette flow.** The fluid motion in the Couette flow apparatus is unidirectional. Therefore, a deviation of a larva's trajectory from a straight line is due to its own motion. The root mean square of the deviation is of the same order of magnitude as the deviation of a larva from the axis of a helix (see Figures 4-5 of larval helical motion).

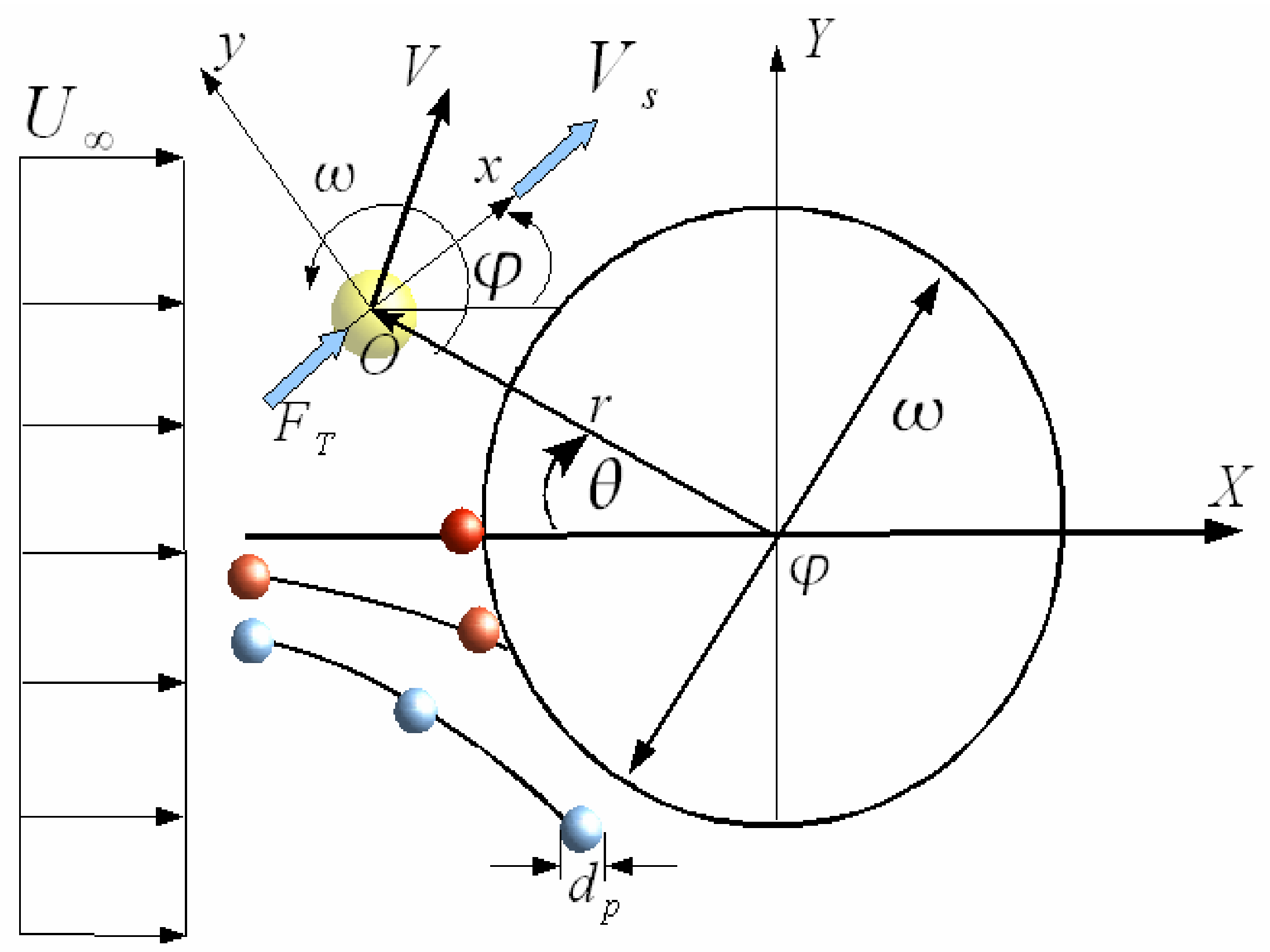

**Figure 7. Orthogonal coordinate systems**. The system *OXY* is fixed in a collector and the system *oxy* is fixed in the larva. The initial *o* of the coordinate system *oxy* coincides with the center of inertia of the larva and moves with linear velocity $\mathbf{V}$, the larva's velocity. The coordinate of the initial $o$ in the coordinate system *OXY* is defined by the radius-vector $\mathbf{r}(X_o,Y_o)$. The coordinate system *oxy* rotates as a whole, with the angular velocity of the larva $\boldsymbol{\omega}$. The angle between the axes *ox* and OX is denoted as $\varphi$. The angle between the radius vector and the negative direction of the longitudinal axis is denoted as $\theta$. The larva moves with the swimming velocity $\mathbf{V}_S$ under the action of a propulsion force $\mathbf{F}_T$.

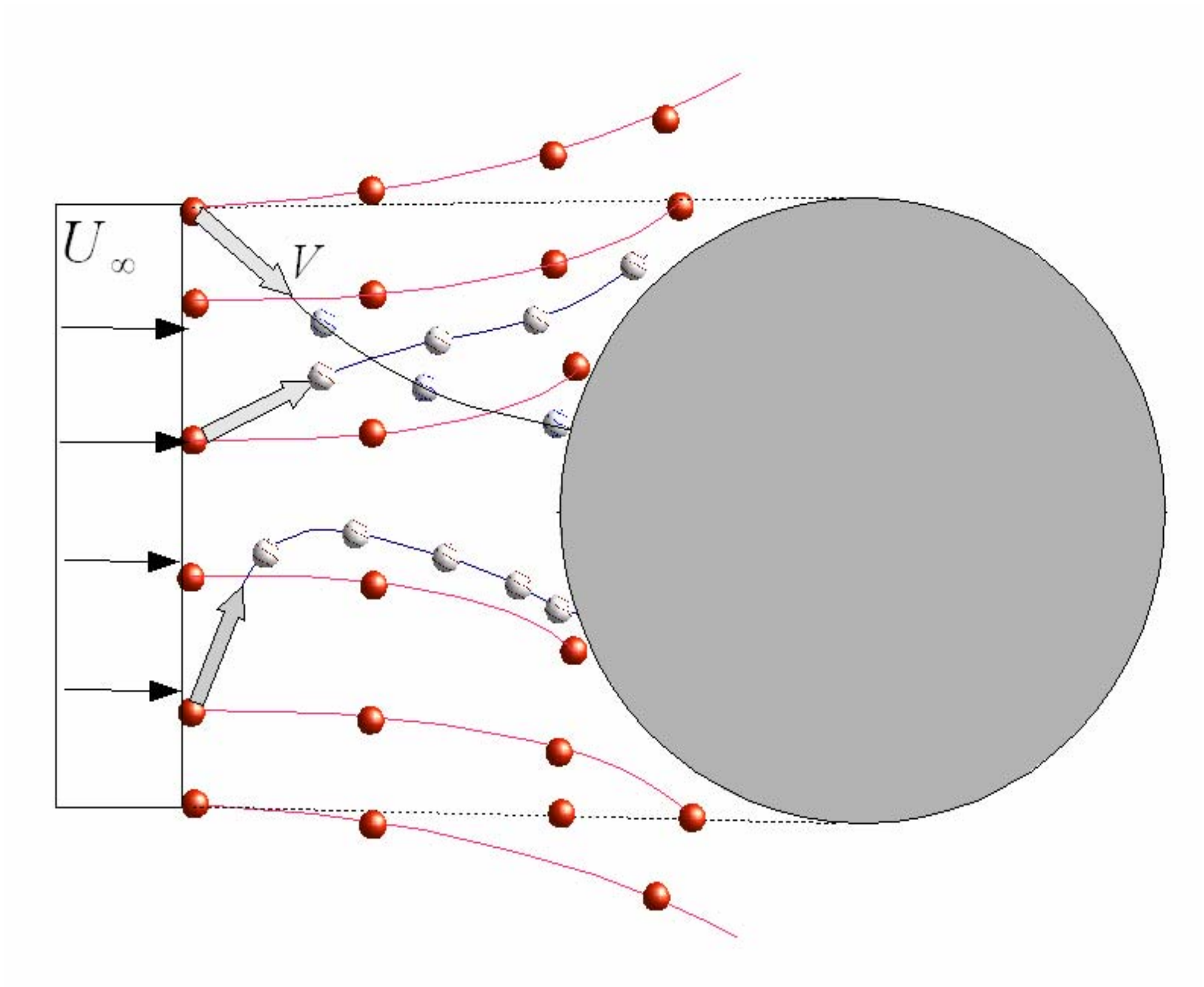

**Figure 8. Schematic trajectories of passive particles (red spheres) and of living larvae (light spheres).** Arrows indicate the initial direction of the vector of self-propulsion.

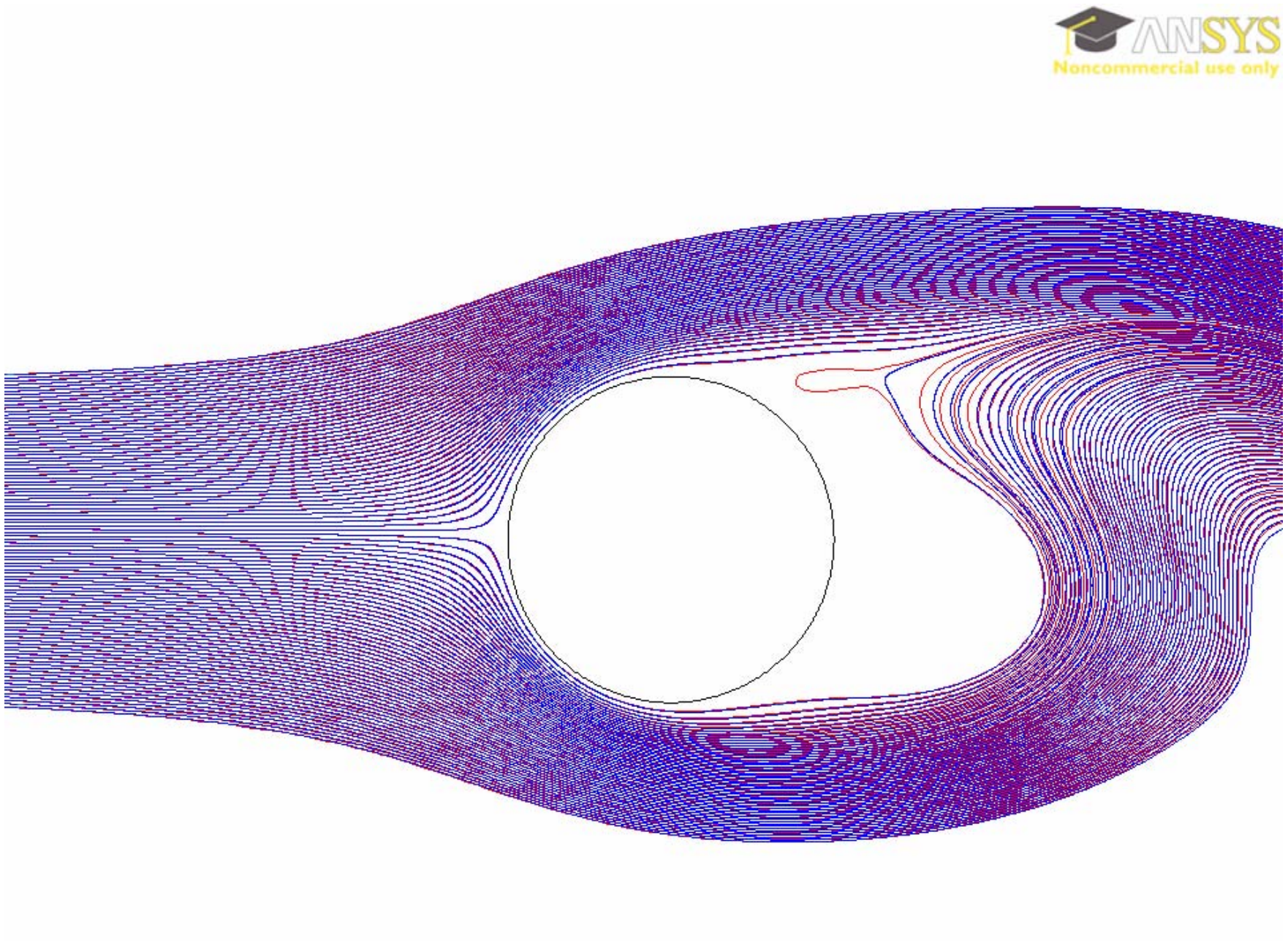

**Figure 9. Streamlines (red) and trajectories of particles (blue) near a cylinder** ($\mathrm{Stk} = 4.4\times10^{-3}$, $\mathrm{Re}_c = 200$).

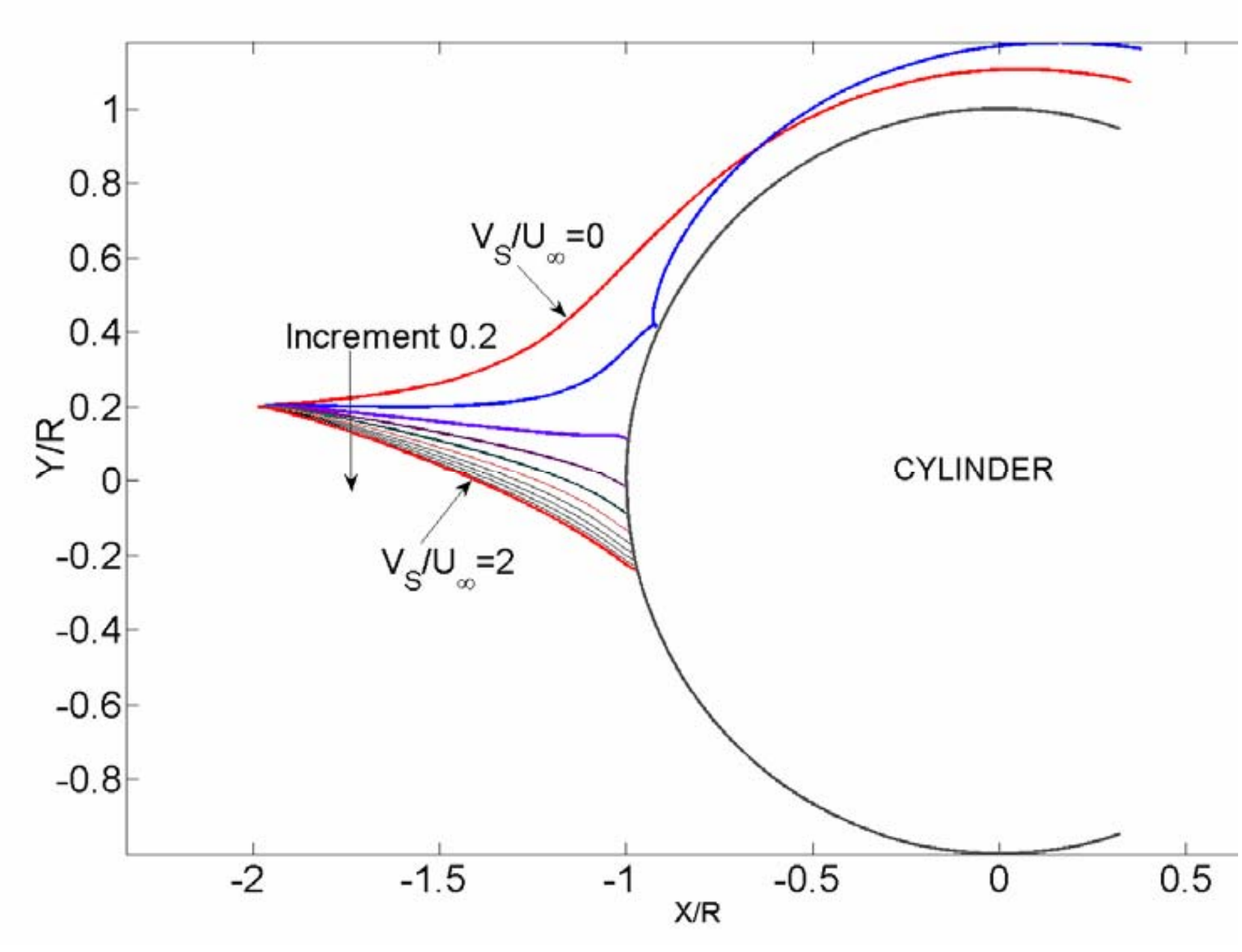

**Figure 10. Trajectories of a self-propelled particle beginning the motion with initial angle of turn $\varphi = -30^{\circ}$ and initial coordinate $Y/R = 0.2$** ($\mathrm{Stk} = 2.2\times10^{-3}$, $\mathrm{Re_c} = 100$). The trajectories can cross or touch the collector depending on the ratio of the swimming velocity of a larva to the velocity of the current and the initial angle of turn. Note that for a passive particle all trajectories are identical to that pertaining to the line $V_S/U_\infty = 0$.

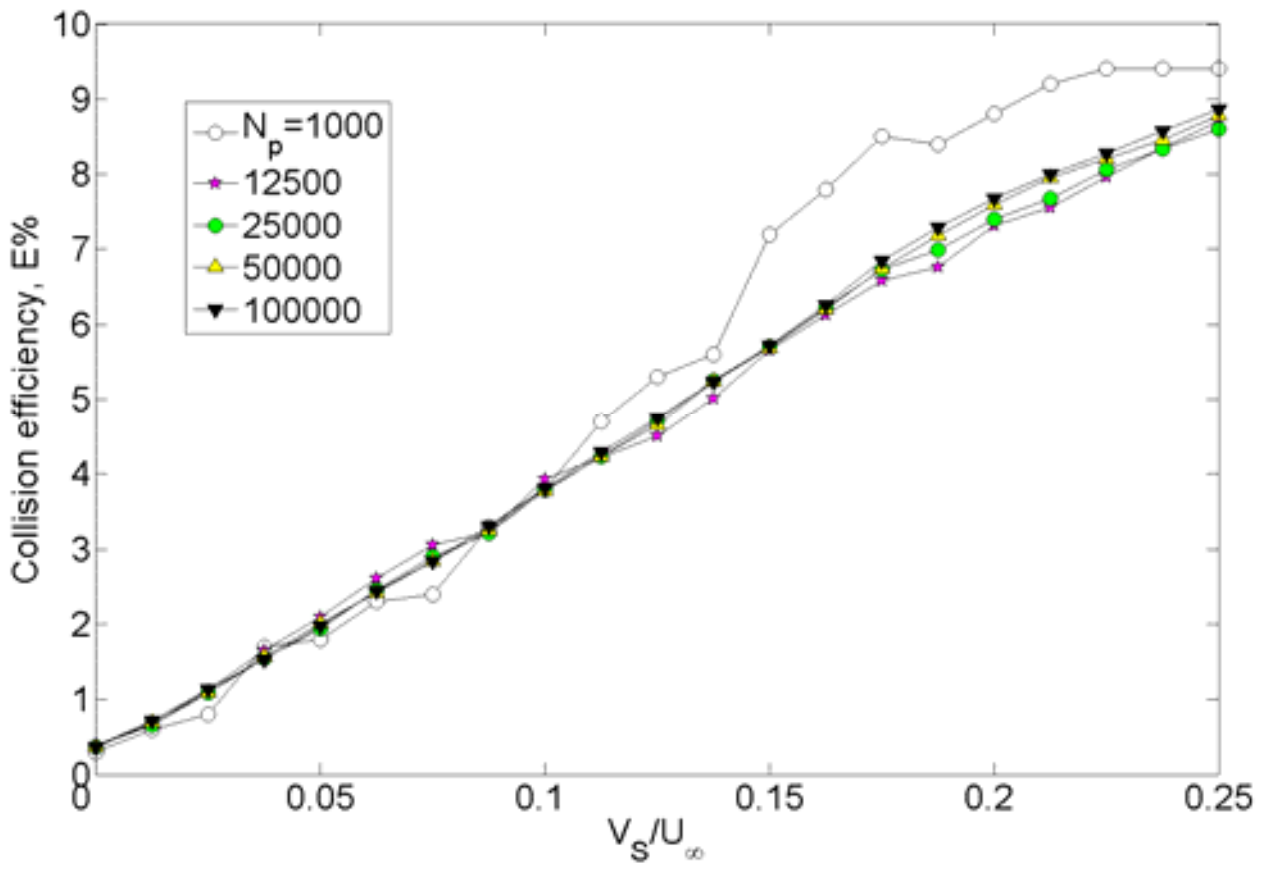

**Figure 11. Collision efficiency as a function of the ratio $V_S/U_\infty$ for a different number of simulated particles $N_p$** ($\mathrm{Stk} = 2.2\times10^{-3}$, $\mathrm{Re_c} = 100$, $\Phi = 45^{\circ}$). To attain a high accuracy of the collision efficiency coefficient, it is necessary to simulate the trajectories of a large number of particles of the order of $N_p = 10^4 - 10^5$.

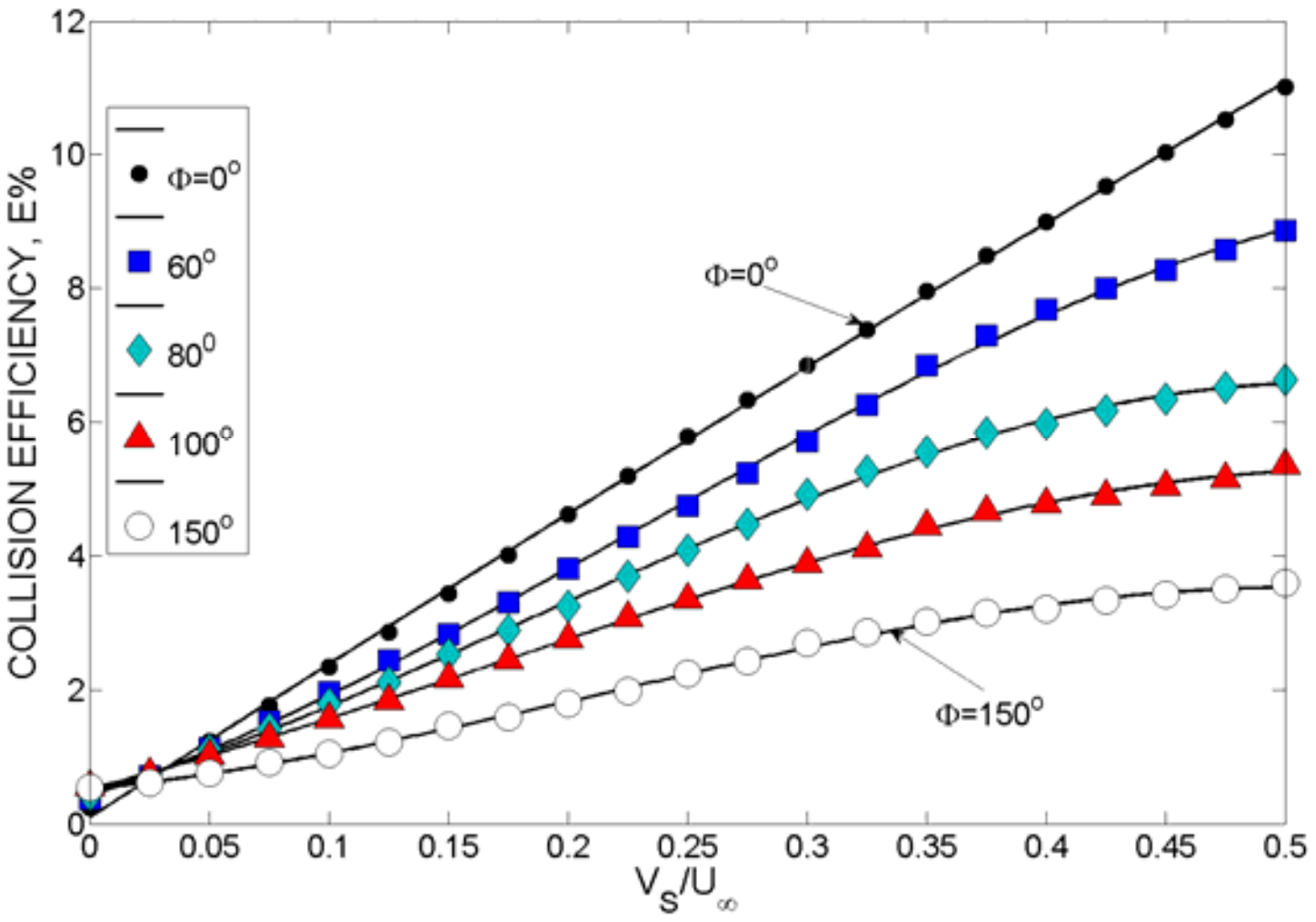

**Figure 12. Collision efficiency as a function of the ratio $V_S/U_\infty$ for different angles Φ ($\mathrm{Stk}=2.2\times10^{-3}$, $\mathrm{Re_c}=100$).** The collision efficiency depends on the preferable direction of motion. For angles $0\le\Phi\le60^\mathrm{o}$, the influence of the initial preferable direction of motion on the collision efficiency is weak; for angles $60^\mathrm{o}\le\Phi\le180^\mathrm{o}$, the influence increases. Further numerical simulations were carried out for $\Phi=60^\mathrm{o}$ when the initial angle of larval direction varies randomly as $-60^\mathrm{o}<\varphi<60^\mathrm{o}$ .

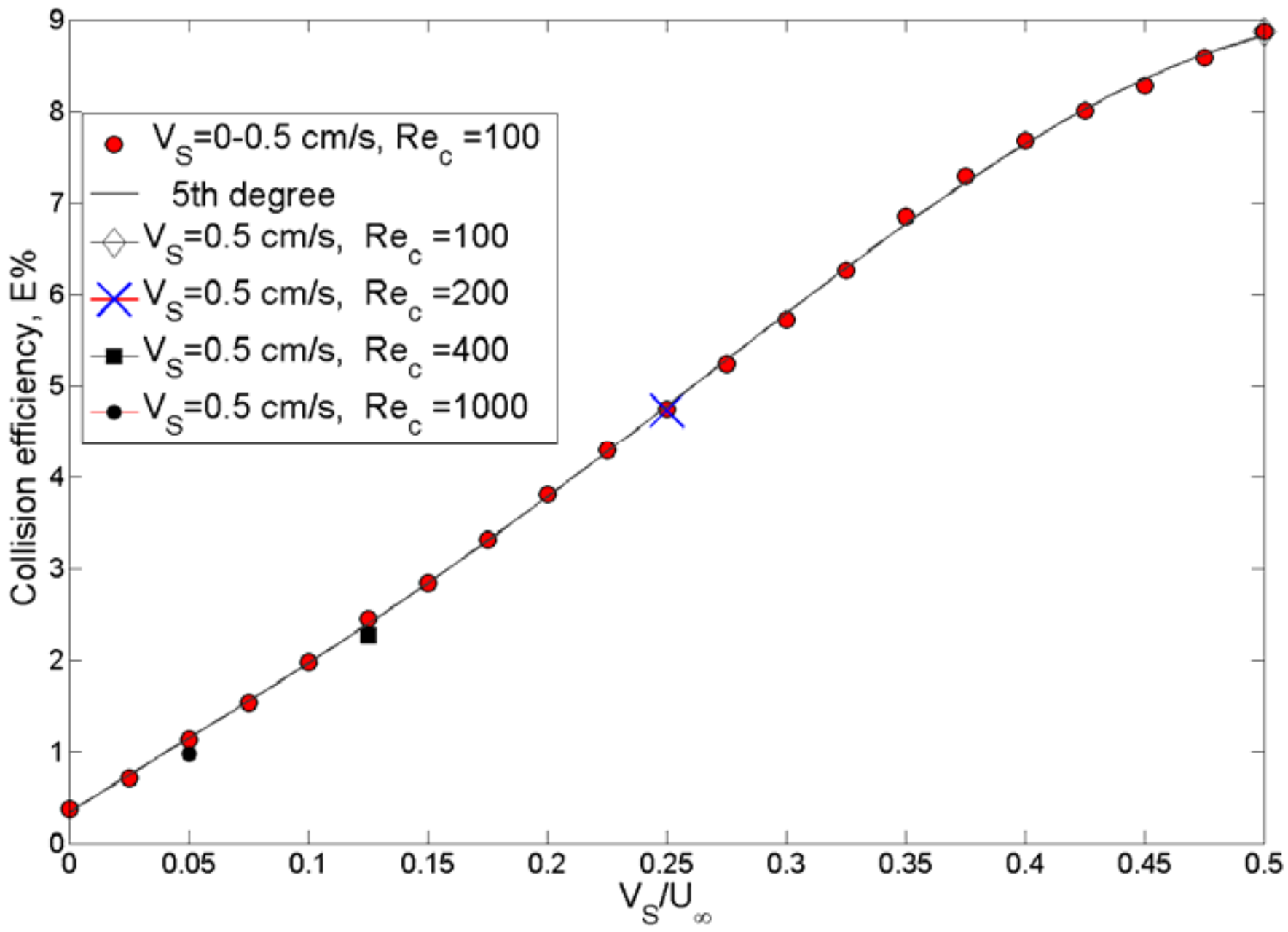

**Figure 13. Collision efficiency as a function of the ratio $V_S/U_\infty$ for $V_S$-const** or **$U_\infty$-const** ($\mathrm{Stk}=2.2\times10^{-3}-2.2\times10^{-2}$).

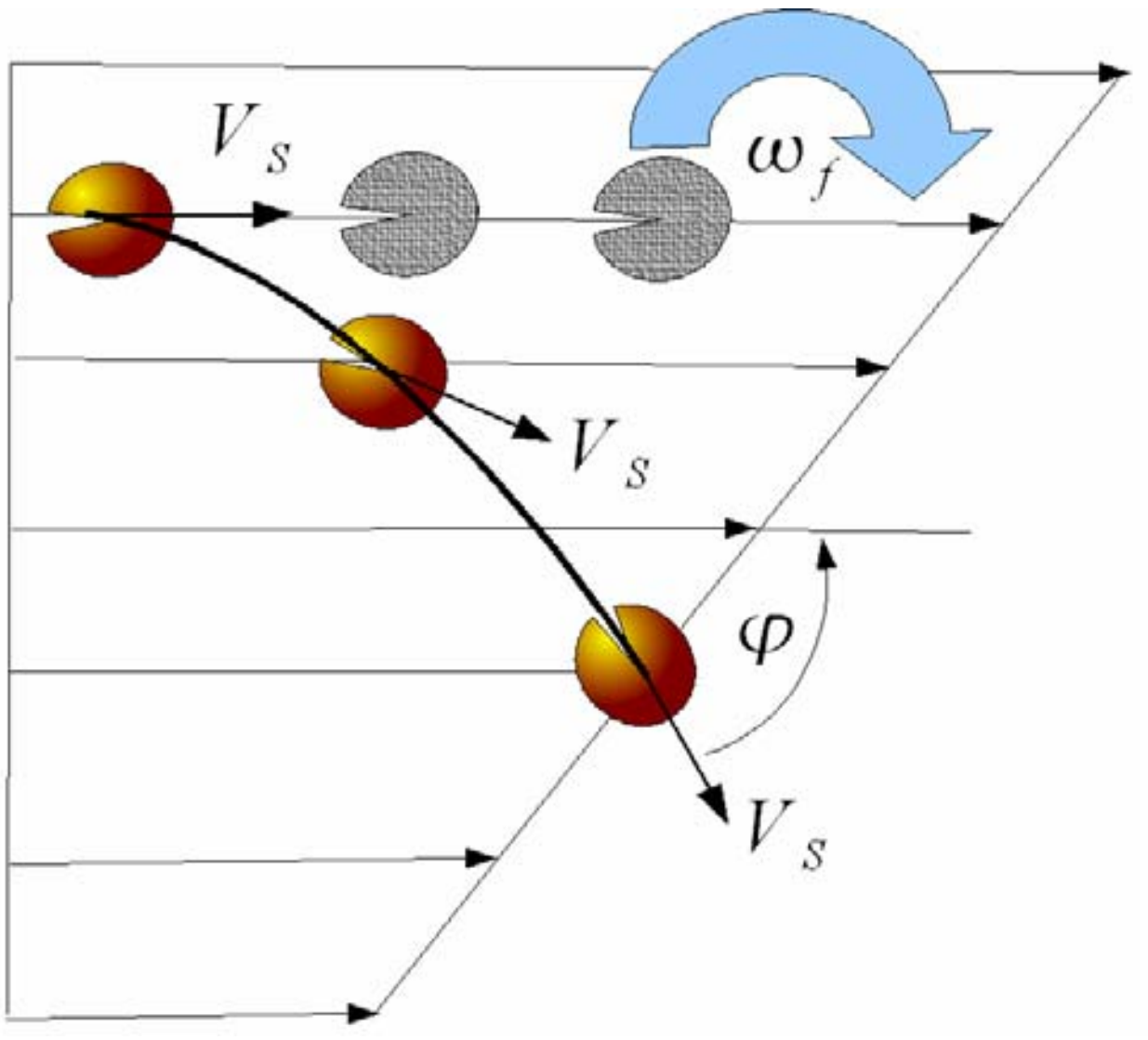

**Figure 14. Motion of small passive and self-propelled particles in viscous shear flow.** Locally the flow in the vicinity of the particles can be viewed as unidirectional. In a unidirectional flow a small passive particle translates with the flow and rotates with an angular velocity $\omega_f$. Within the framework of the low Reynolds number hydrodynamics the center of a passive rotating particle moves rectilinearly. For a self-propelled particle the angle $\varphi$ between the vector of the swimming velocity $\mathbf{V}_S$ and the particles direction of motion varies with time. The sum of the two velocity vectors $\mathbf{U}+\mathbf{V}_S$ forces the particle to move along a curvilinear trajectory instead to continue a rectilinear motion.

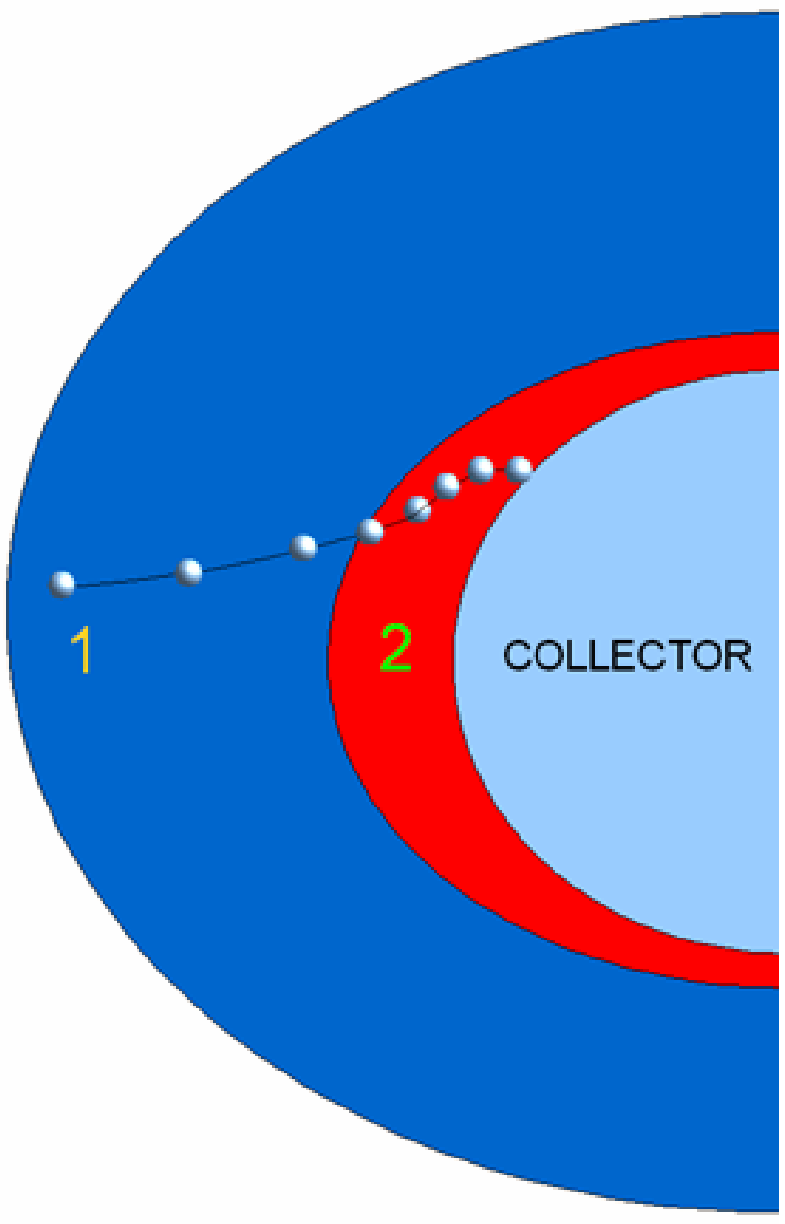

**Figure 15. Fluid domains of influence of a collector on a larva**. 1. Area of influence of a collector on a larva where the fluid viscosity can be neglected; 2. The boundary layer of the collector where the fluid viscosity have to be taken into account.

# Tables

**Table 1. Characteristic range of the parameters for small-scale, wind driven oceanic turbulence**

| $L_\varepsilon$ | $\varepsilon$ | $\eta_k$ | $u_k$ | $T_k$ | $\mathrm{Re}_\lambda$ |
|---|---|---|---|---|---|
| 2-100 m | $10^{-4}-10^{-1}\mathrm{cm}^2/\mathrm{s}^3$ | 0.3-2 mm | $0.3-3$ mm/s | 5-0.1 s | $200-10^4$ |

**Table 2. Parameters of sea turbulence and the collision efficiency of a cylinder collecting passive particles calculated using the $k-\omega$ model.**

| $\bar{U}_\infty$ | $D_c$ | $\mathrm{Re}_c$ | $d_p$ | $L_\varepsilon$ | $\eta_k$ | $\varepsilon$ | $I$ | $\mathrm{Re}_\lambda$ | $E$ |
|---|---|---|---|---|---|---|---|---|---|
| cm/s | cm | - | $\mu$m | m | mm | $\mathrm{cm}^2/\mathrm{s}^3$ | % | - | - |
| 5 | 1 | 500 | 200 | 2.2 | 1.5 | $2\times10^{-3}$ | 23 | 500 | 0.67% |